\documentclass[sigconf]{acmart}

\AtBeginDocument{%
  \providecommand\BibTeX{{%
    \normalfont B\kern-0.5em{\scshape i\kern-0.25em b}\kern-0.8em\TeX}}}

\usepackage{enumitem}
\setlist[itemize]{leftmargin=4mm}

\begin{document}

\title[Immersive Insights]{Immersive Insights: A Hybrid Analytics System for\\Collaborative Exploratory Data Analysis}


\author{Marco Cavallo}
\affiliation{%
  \institution{IBM Research}
}
%\email{marco@mastercava.com}

\author{Mishal Dholakia}
\affiliation{%
  \institution{IBM Research}
}

\author{Matous Havlena}
\affiliation{%
  \institution{IBM Research}
}

\author{Kenneth Ocheltree}
\affiliation{%
  \institution{IBM Research}
}

\author{Mark Podlaseck}
\affiliation{%
  \institution{IBM Research}
}

\renewcommand{\shortauthors}{Cavallo, et al.}

\setcopyright{acmcopyright}
\copyrightyear{2019} 
\acmYear{2019} 
\acmConference[VRST '19]{25th ACM Symposium on Virtual Reality Software and Technology}{November 12--15, 2019}{Parramatta, NSW, Australia}
\acmBooktitle{25th ACM Symposium on Virtual Reality Software and Technology (VRST '19), November 12--15, 2019, Parramatta, NSW, Australia}
\acmPrice{15.00}
\acmDOI{10.1145/3359996.3364242}
\acmISBN{978-1-4503-7001-1/19/11}

\begin{abstract}
	% The rise of AR/VR and data science
In the past few years, augmented reality (AR) and virtual reality (VR) technologies have experienced terrific improvements in both accessibility and hardware capabilities, encouraging the application of these devices across various domains.
While researchers have demonstrated the possible advantages of AR and VR for certain data science tasks, it is still unclear how these technologies would perform in the context of \textit{exploratory data analysis} (EDA) at large.
In particular, we believe it is important to better understand which level of immersion EDA would concretely benefit from, and to quantify the contribution of AR and VR with respect to standard analysis workflows.

%Dataspace and Immersive insights
In this work, we leverage a Dataspace reconfigurable hybrid reality environment to study how data scientists might perform EDA in a co-located, collaborative context.
Specifically, we propose the design and implementation of \textit{Immersive Insights}, a hybrid analytics system combining high-resolution displays, table projections, and augmented reality (AR) visualizations of the data.

% Evaluation
We conducted a two-part user study with twelve data scientists, in which we evaluated how different levels of data immersion affect the EDA process and compared the performance of Immersive Insights with a state-of-the-art, non-immersive data analysis system.

\end{abstract}

\begin{CCSXML}
<ccs2012>
 <concept>
  <concept_id>10010520.10010553.10010562</concept_id>
  <concept_desc>Computer systems organization~Embedded systems</concept_desc>
  <concept_significance>500</concept_significance>
 </concept>
 <concept>
  <concept_id>10010520.10010575.10010755</concept_id>
  <concept_desc>Computer systems organization~Redundancy</concept_desc>
  <concept_significance>300</concept_significance>
 </concept>
 <concept>
  <concept_id>10010520.10010553.10010554</concept_id>
  <concept_desc>Computer systems organization~Robotics</concept_desc>
  <concept_significance>100</concept_significance>
 </concept>
 <concept>
  <concept_id>10003033.10003083.10003095</concept_id>
  <concept_desc>Networks~Network reliability</concept_desc>
  <concept_significance>100</concept_significance>
 </concept>
</ccs2012>
\end{CCSXML}

\ccsdesc[500]{Human-centered computing~Visualization Systems and Tools}
%\ccsdesc[500]{Human-centered computing~Interactive systems and tools}
\ccsdesc[500]{Human-centered computing~Mixed / Augmented Reality}

\keywords{Dataspace, Hybrid Reality, Data Visualization, Exploratory Data Analysis, Augmented Reality, Virtuality Continuum, Clustering}

\begin{teaserfigure}
	\centering
  \includegraphics[width=0.95\textwidth]{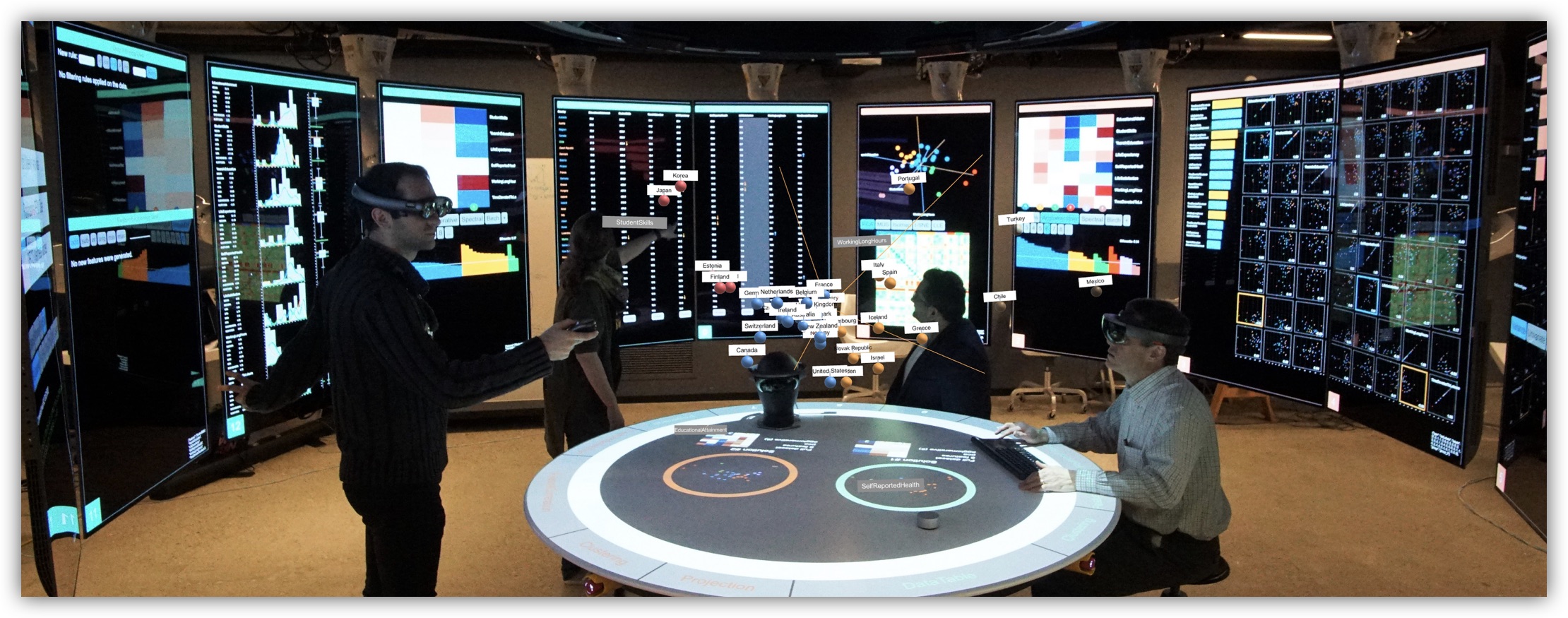}
  \caption{Immersive Insights is a collaborative, hybrid analytics system for exploratory data analysis.
 %, focusing on projection, clustering and statistical analysis. 
 The application, implemented in a Dataspace environment \cite{cavallo2019dataspace}, leverages multimodal interaction and combines the use of technologies such as 1) high-resolution movable screens to visualize statistical information, 2) a central projection table providing an overview of the current analysis, and 3) an augmented reality view to visualize and interact with high-dimensional data. 
%Immersive Insights enables data scientists to simultaneously analyze independent parameterized subsets of the data, called ``data instances'' or ``solutions''.
}
  \label{fig:teaser}
\end{teaserfigure}

\maketitle

\section{Introduction}
Exploratory Data Analysis (EDA) is the process of performing initial inquiries on a dataset, with the goal of discovering interesting patterns, spotting anomalies, testing hypotheses and checking assumptions \cite{tukey1977exploratory}.
In order to make sense of data, data scientists combine many graphical and numerical techniques, drawing elements from statistical inference, clustering analysis, dimensionality reduction, and sensitivity analysis. In this sense, there is no standard workflow for performing EDA, which, in its current form, vastly varies based on the type of data and on the assumptions that an analyst wants to assess.
%[C], and becomes complex to manage in collaborative settings.
%Examples of what already exists
Scripting frameworks (e.g. R, Matlab) and data visualization tools have been developed to assist data scientists during the EDA process, but are mostly desktop-based and do not provide support for collaborative analysis.
% Immersive analytics
Over time, researchers have repeatedly tried to propose immersive technologies (such as CAVE systems \cite{cruz1992cave,cruz1993surround,defanti2011future,manjrekar2014cave}, powerwalls \cite{marrinan2014sage2,marai2016interdisciplinary}, and virtual and augmented reality headsets) as a possible solution for specific data science tasks, defining a research area known as \textit{Immersive Analytics}  \cite{chandler2015immersive,marriot2018immersive}.
% Recent renewed interest
The technological improvements and hardware commoditization of head-mounted-displays (HMDs) of the past few years have reignited interest in applying these devices to domains such as EDA, and a few companies have even proposed some first VR-based commercial solutions \cite{3data,virtualitics}.
% Limitations
Despite many successful research demos and various marketing campaigns showcasing the advantages of full immersion and natural interaction with 3D data, no standalone AR/VR software tool---to the best of the authors' knowledge---has been able to contend with the complexity of collaborative EDA at large and therefore compare favorably with existing desktop-based software tools.
%Notes to take inspiration from
%none of the existing immersive analytics research prototypes have grown complex enough to allow a fair comparison with existing laptop-based visual analytics tools, mostly due to the convoluted requirements of collaborative EDA.
%AR and VR technologies still present limitations in resolution, field of view, comfort and co-presence \cite{cavallo2019dataspace}.
%EDA systems that could provide a proper have been proposed yet, mostly due to the complex requirements of EDA (e.g. variety of analytical tasks, different data granularities, focus-and-context capability, data instances comparison).

% Our proposal
In this work, we take a step back and ask ourselves which level of immersion 
%[C](along the \textit{virtuality continuum} \cite{milgram1994taxonomy}) 
a fully-fledged exploratory data analysis session would actually benefit from.
To delve into this, we leveraged a Dataspace \cite{cavallo2019dataspace} hybrid reality system, a reconfigurable room-scale environment that builds on the legacy of CAVE2 and other smart workspace concepts \cite{Raskar:1998:OFU:280814.280861,Prante:2004:RCD:1038051.1038085,Jones:2014:RME:2642918.2647383,farrell2016symbiotic, lischke2018interacting}, and is characterized by unique \textit{flexible data immersion} capabilities. 
%[C]and seamless integration with AR and VR devices.
% Immersive insights
In particular, we introduce \textit{Immersive Insights}, a \textit{hybrid analytics} Dataspace application for collaborative exploratory data analysis. Immersive Insights attempts to cover a broad range of EDA tasks (including statistical analysis, dimensionality reduction and clustering) in the context of tabular data analysis, freely taking inspiration from the state-of-the-art, desktop-based clustering tool \textit{Clustrophile 2} \cite{cavallo2018clustrophile}.

%User study
We evaluated Immersive Insights through a two-part user study involving twelve data scientists. In the first part, participants performed a set of EDA tasks using Immersive Insights in different modalities (the Dataspace environment only, AR integration, AR standalone, VR standalone), allowing us to time their performance at different levels of immersion.
The second part of the study compares Immersive Insights with its desktop-based counterpart Clustrophile 2, attempting to quantify the eventual performance improvement achieved through immersive technologies during a more complex EDA session.

%Outline the paper
In the following, we first review the literature on data visualization for EDA, from standard desktop-based tools to immersive environments and AR/VR headsets. Then we briefly describe the Dataspace environment, and articulate the rationale underlying our Immersive Insights implementation. We then describe our user study and its results, concluding with a few considerations regarding the integration of new technologies into EDA and the future of collaborative data analysis.

\section{Related Work}

We have built our work upon existing software, and the extant literature on EDA, specifically those tools and texts concerning dimensionality reduction and clustering analysis with statistical support. We first consider how state-of-the-art desktop-based tools are currently being used for EDA, and then discuss how researchers have previously attempted to employ immersive environments such as CAVEs, tiled-display walls and head mounted displays (HMDs) for visual data exploration and analysis.

\subsection{Tools for Exploratory Data Analysis}

\subsubsection{Visualizing High-dimensional Data.} Dimensionality reduction (DR) is the process of reducing total features (variables, dimensions) in the data, for purposes of preprocessing or visualization.
%Feature projection (feature extraction), in particular, reduces an initial set of features to a smaller set of derived features intended to be descriptive of the original dataset, and non-redundant.
For example, algorithms such as PCA \cite{tipping1999probabilistic} are often used to reduce total dimensions to two, so as to visualize high-dimensional data on a 2D plane (e.g. in the form of a scatterplot) and therefore interpret similarities among data points and possible structures.
%Start with the related work
Due to the lack of a clear mapping between the axes generated by DR techniques and the original data dimensions, earlier work in data analysis has proposed various tools and techniques to guide users in exploring low-dimensional projections of data \cite{liu2017visualizing,engel2012survey}.
%Rotations
Methods such as rotation and isolation \cite{PRIM9_1974} enable the user to interactively rotate multivariate data, and statistical information can be used to structure possible visualizations of the data \cite{Asimov_1985,seo:infovis04,Wills_2008,Wongsuphasawat_2016,Vartak_2015a,Demiralp:2017:VLDB,cavallo2018clustrophile}.
%Visualization in DR Scatter Plots
Since low-dimensional projections are generally lossy representations of high-dimensional data relations, researchers have introduced visual methods to convey and correct dimensionality reduction errors \cite{aupetit2007visualizing, lespinats2011checkviz, chuang2012interpretation, stahnke2016probing}. 
% Feature contribution
Similarly, enhanced biplots \cite{coimbra2016explaining,gabriel1971biplot} and \textit{prolines} \cite{cavallo2018visual,faust2019dimreader} have been introduced to visualize the contribution of data features to the DR plane.
%Direct Manipulation in DR
Researchers have also used direct manipulation to interactively modify data
through DR visualizations \cite{orban2019drag,jeong2009ipca} and out-of-sample extrapolation \cite{bengio2004out,van2009dimensionality}. A recent what-if-analysis example is represented by the forward and backward projection interaction introduced by Praxis \cite{cavallo2018visual}.
%This allows for iterative what-if analysis without re-running dimensionality reduction algorithms.

\subsubsection{Identifying Structures in the Data.} Clustering is the task of grouping sets of objects so that members of the same group (``cluster'') are more similar to each other than to those in other groups, according to some specific distance measure. Combined with statistical analysis, clustering is often used in EDA to discover and characterize salient structures in a dataset \cite{xu2015comprehensive,jain1999data}.
%Start
To improve user understanding of clustering results across domains, early interactive systems introduced the use of coordinated visualizations with drill-down/up capabilities \cite{Jinwook_Seo_2002} and the visual comparison of different clustering results \cite{Lex_2010,Cao_2011,Lyi_2015,Pilhofer_2012,Jinwook_Seo_2002}.
%Combination
To contextualize the various assignments generated by clustering algorithms, tools such as ClustVis~\cite{metsalu2015clustvis}, Clustrophile~\cite{demiralp2016clustrophile} and ClusterVision~\cite{kwon2018clustervision} coordinate visualizations of discrete clusterings with scatterplot visualizations of dimensionality reductions. Correlation and ANOVA-based significance analyses are seamlessly integrated in the clustering process in Clustrophile 2 \cite{cavallo2018clustrophile}.\\
%This tool combined these analyses with distributional information to obtain interpretable cluster descriptions and a new task-driven type of analysis, focusing less on numerical indices and more on the usefulness of a clustering result with respect to the user's interests.\\
%%Close the section
Immersive Insights takes inspiration from Clustrophile 2's combination of data projections, clustering results, distributional information and significance testing, and complements them with aggregate views to facilitate collaborative analysis. 
%[C]In particular, Immersive Insights integrates forward and backward projections and prolines in a dedicated AR view, and supports collaboratienables multimodal interaction in collaborative settings.
%[C] where various visualization layouts can be dynamically applied to independent data instances, so that analysts can work in parallel on different subsets of the data and then share their results.

\subsection{Immersive Data Visualization}
%Limits of desktop-based applications
\subsubsection{Virtual Reality Theaters and Tiled Display Walls.} Due to the continuous growth in size and complexity of data digitally collected and stored, researchers have tried to find ways to make sense of these datasets through the development of new visualization instruments. While desktop-based applications can provide overview information through summarization, abstraction, and focus-plus-context techniques, the necessity of fitting vast quantities of data on a single display is often detrimental to multi-scale data exploration \cite{ni2006increased}.
%Environments for data visualization
By surrounding users with visuals through ``immersion,'' physical data visualization systems have been effective in allowing users to explore 3D spatial data (such as molecules, astrophysical phenomena, and geoscience datasets) and high-dimensional data in general \cite{mcintire2014possible}.
%The split: CAVEs vs powerwalls
%Since the '90s, immersive virtual environments such as CAVE \cite{cruz1993surround} and high-resolution tiled display walls (also known as ``powerwalls'') have represented two distinct solutions to the large-scale data visualization problem.
%1: CAVEs
CAVE (CAVE Automatic Virtual Environment), introduced in 1992, was a cube measuring 10 feet on each side, and utilized a set of projectors to allow a small number of researchers to experience stereoscopic 3D graphics on five of its sides.
%2: Display walls
Leveraging improvements in LCD technologies made in the early 2000s, tiled display walls (also known as ``powerwalls'') gradually arose as a viable alternative to CAVE systems, offering superior image quality and resolution with relatively low maintenance. Despite renouncing stereoscopic 3D rendering, display walls still enabled the visualization of large datasets, and provided both detail and context. This shift opened up new possibilities for collaborative data analysis, as demonstrated by the use of SAGE \cite{renambot2004sage} (and later SAGE2 \cite{marrinan2014sage2}) in the EVL Cybercommons room \cite{marai2016interdisciplinary, krumbholz2005lambda}.
%Recent years
In recent years, researchers have explored a number of ways to further improve on these two technologies \cite{defanti2011future,manjrekar2014cave}, by sensibly increasing resolution \cite{papadopoulos2015reality}, providing more flexibility in screen configuration \cite{ponto2015dscvr}, integrating mobile devices \cite{krum2014tablet,horak2018david}, and exploring interaction with artificial agents \cite{farrell2016symbiotic, venkataraman2016ceding}.
%Putting it all together
CAVE2 \cite{febretti2013cave2}, a system composed of 72 cylindrically-positioned displays, aimed at combining the effectiveness of CAVE systems in visualizing 3D datasets with the capabilities of more recent ultra-high-resolution environments, which were a better fit for 2D data visualization. By combining the SAGE \cite{renambot2004sage} tiled display system and OmegaLib \cite{febretti2014omegalib} virtual reality middleware, CAVE2 enabled researchers to seamlessly interact with large collections of 2D and 3D data, providing the first full implementation of a Hybrid Reality Environment (HRE) \cite{febretti2013cave2}.

%Switch to HMDs!
\subsubsection{HMD-based Immersive Analytics.} While CAVE was originally introduced as an alternative to existing bulky desk-based head-mounted displays (HMDs) \cite{sutherland1965ultimate}, recent technological advances have drastically improved the resolution, field of view, form factor and availability of virtual and augmented reality headsets, which have once again become tractable tools for data visualization \cite{marriot2018immersive,chandler2015immersive}.
% Limitations of Immersive environments
In fact, HMD-based Immersive Analytics \cite{cliquet2017towards} provide an effective solution to the high cost, complex maintenance, and scalability limits of immersive environments \cite{papadopoulos2015scalability}.
%Examples
%HMDs have successfully been applied to the visualization of brain information \cite{he2017medical} and other scientific data, and have been used in a variety of data visualization settings, including smartphone-based VR \cite{butcher2016immersive}, desk VR applications \cite{zielasko2017buenosdias}, and marker-based AR \cite{ritsos2017synthetic}.
%Prior art validating their use
Millais et al. \cite{millais2018exploring} demonstrate the advantages of using immersion for data exploration in virtual reality, while Butscher et al. \cite{butscher2018clusters} examine how immersive technologies can facilitate collaborative analysis to better detect clusters, trends and outliers. However, as outlined by McIntire et al. \cite{mcintire2014possible}, the use of stereoscopic displays alone for information visualization still has limitations. While HMDs may be convenient in performing tasks associated with spatial or multidimensional data, they can fall short in displaying statistical and abstract information, which is instead more successfully handled by 2D visualizations \cite{marriot2018immersive,bach2016immersive}. \\
% DATASPACE!
Dataspace \cite{cavallo2019dataspace} attempts to bring together the advantages of all the workspaces outlined above, combining high-resolution displays with augmented reality headsets, plus a central interactive table and integrated AI-based cognitive functionalities. Using Dataspace, Immersive Insights enables collaborative analysis of spatial datasets by allowing users to rapidly move between high-resolution statistical information (displayed on 2D screens) and 3D representations of high-dimensional data (visualized in AR).

\section{Dataspace}

\begin{figure}
\includegraphics[width=\columnwidth]{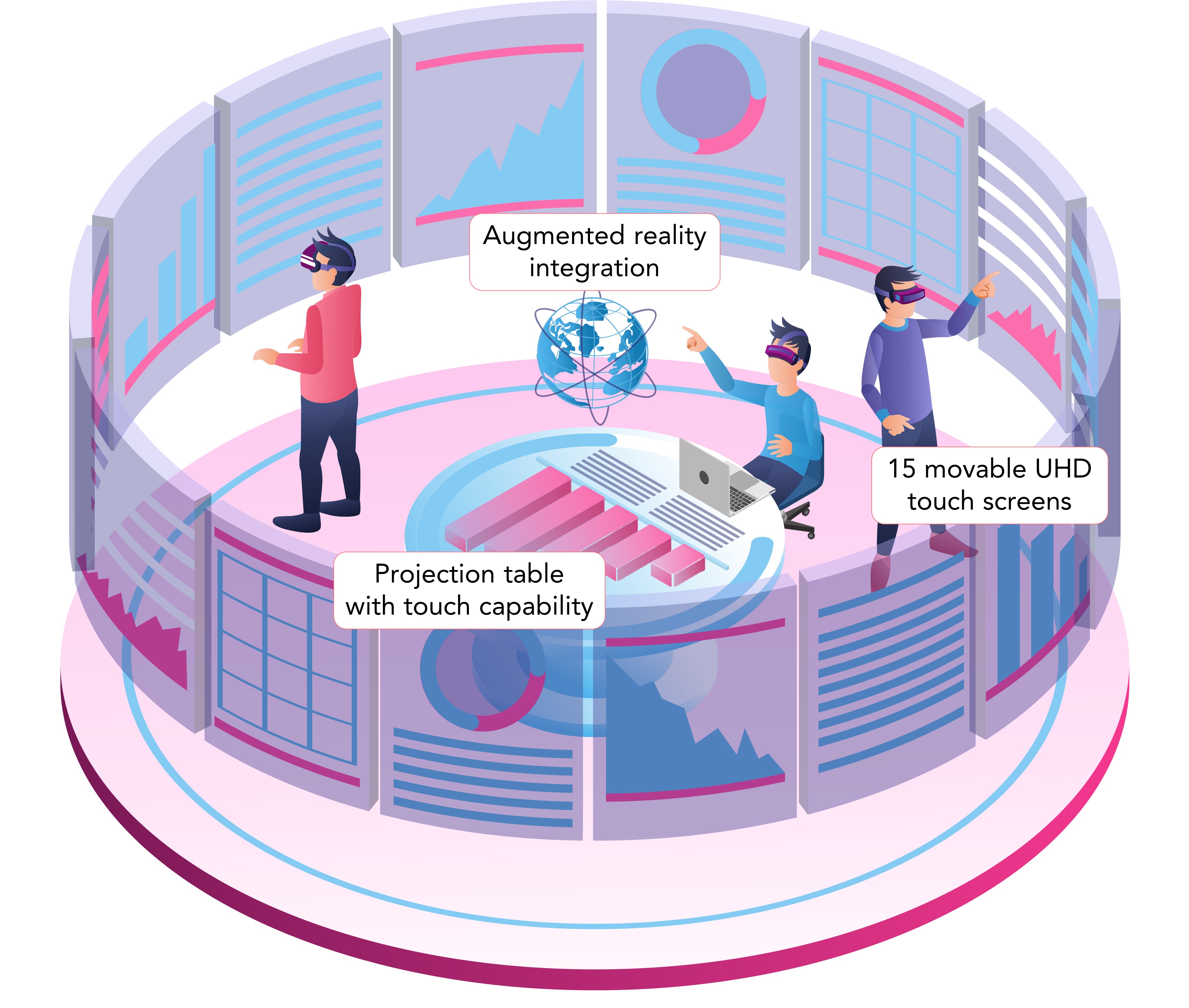}
\caption{Dataspace is an immersive, collaborative, and reconfigurable hybrid reality environment, combining heterogeneous technologies and mixed interaction methodologies \cite{cavallo2019dataspace}. The workspace includes 15 large high-resolution displays attached to moving robotic arms, two table projectors, and integrates AR and VR headsets, laptops and other mobile devices.}
\label{fig:dataspace}
\vspace{-1.5em}
\end{figure}

Dataspace (Fig.~\ref{fig:dataspace}) is a room-sized \textit{hybrid reality environment}  where people can work together and interact naturally with both 2D and 3D information. Dataspace leverages a unique combination of movable high-resolution displays, an interactive projection table, and augmented and virtual reality head-mounted displays, and is specifically aimed at improving the collaborative decision-making process. In this section we briefly outline the main characteristics of the environment to contextualize our Immersive Insights system. We refer readers to the Dataspace paper \cite{cavallo2019dataspace} for further details.

\subsection{The Environment}
The primary physical components constituting Dataspace are:
\begin{itemize}
\item 15 UHD displays with touch capability, which can be moved and rotated in space via robotic arms mounted to the ceiling.
\item A central table onto which visual content can be displayed through two HD projectors. Touch and gestures performed on the table are detected through a set of eight Kinect v2 sensors.
\item A spatial audio system consisting of 20+2 speakers, and an array of four directional microphones that can be used to detect voice commands and their sources.
\item A set of augmented reality headsets (currently Microsoft Hololens and Magic Leap One devices) to interact with spatial or high-dimensional data, often visualized atop the central table.
\item A set of virtual reality headsets (Samsung Odyssey) to remotely access the environment and its functionalities, providing a virtual replica of Dataspace and its content.
\end{itemize}

Dataspace implements a modular software architecture for 1) safely coordinating robotic arms, 2) detecting and tracking people and objects in the environment, 3) controlling screen content and table projections, 4) interpreting speech and generating audio output.
The environment adopts web-based rendering through Electron \cite{electron}, promoting high flexibility for application development (e.g. HTML, WebGL) and support for external devices with browsing capabilities (e.g. laptops and smartphones).
Thanks to the combination of UHD screens and AR headsets, Dataspace can  \textit{simultaneously render high-resolution 2D content and 3D information}, and may thus be referred to as a \textit{hybrid reality} environment \cite{febretti2013cave2}. Dataspace can provide \textit{flexible data immersion} through its capability to dynamically reconfigure screens in space and through seamless integration with AR and VR devices.
Thanks to its \textit{spatial awareness system} feature, Dataspace can leverage the relative positioning of different environmental elements, enabling a wide set of context-aware interactions. Examples include automatically performing actions on the screen(s) closest to a specific user, orienting table content and lighting towards the person who is currently speaking, physically moving information closer to the user who just performed a gesture, and moving content to different screens through gaze or AR/VR controllers.

\section{Immersive Insights}
Immersive Insights is a web-based system specifically developed for Dataspace, and aimed at enhancing the EDA experience of a group of data scientists who want to make sense of pre-existing, often unlabeled, datasets. The current version of Immersive Insights focuses on statistical analysis, clustering, dimensionality reduction, and feature sensitivity analysis.
%, extending visualization and interaction methods recently proposed in the field and applying them through a combination of heterogeneous technologies.
%Outline this section
In this section we first outline the design challenges characteristic to EDA, and then describe the data flow, visualizations and interactions associated with the Dataspace screens, table, and augmented reality extension.
\paragraph{System architecture.} Immersive Insights' front end has been implemented with React.js \cite{react} in combination with D3.js \cite{d3}, and communicates through websockets with a back end implemented in Python. The back end takes care of computational tasks by leveraging standard libraries such as numpy, scikit-learn and pandas, and forwards the results to associated views.
Immersive Insights can be run in development mode on a normal laptop and tested in Dataspace, or be directly deployed to the environment as Docker container. We note that our system does not necessarily require Dataspace to work, and can be accessed from any web browser.
%[C] (though doing so sacrifices certain types of interaction and computational power).

\subsection{Design Considerations}

\subsubsection{Getting lost in EDA}

Identifying relevant structures in and having insights about unlabeled data is a non-trivial process in which data scientists iteratively apply a varying set of algorithms and statistical methods. There is, by definition, no standard way of performing EDA, and steps such as data preprocessing, dimensionality reduction, clustering, and formulation and validation of hypotheses about the data are continuously and variously used by each analyst.
The absence of a standardized workflow, combined with the wide variety of data analysis tasks and algorithms, poses an initial design challenge involving the reproducibility of EDA sessions. This problem is exacerbated by the \textit{multiscale nature} of data exploration \cite{cavallo2018clustrophile}: identifying relevant insights often depends upon the exploration of many different subsets of the original dataset. Techniques such as filtering, isolation, and sub-clustering are commonly used to identify and better understand smaller structures in the data. Similarly, feature selection involves considering subsets not of data samples, but of data dimensions (features), in order to study their statistical significance.
When building software tools to support EDA, it is inherently difficult to keep track of all these iterative data transformations, drill-down operations, and the ways different techniques affect various subsets of the data. This is also the reason why most general-purpose EDA tools do not support collaborative workflows.\\
A second significant challenge is providing a \textit{contextualized understanding} of how algorithms apply differently to the data and its various substructures \cite{tukey1977exploratory}. This generally involves making statistical information available to describe the data subset currently under consideration, and the ability to visually and quantitatively compare that data subset with other subsets, or the original dataset. Standard data visualization techniques such as linked views, focus-and-context, and modal windows are generally used in EDA tools to provide contextualization and comparison capabilities. However, screen size and resolution can limit the amount of contextual information that can be simultaneously displayed, often requiring users to navigate too many tabs or click too many buttons. %before obtaining the desired information

\subsubsection{The promise of hybrid reality}
We propose Immersive Insights as a possible solution to the aforementioned limitations of EDA:
workflow reproducibility, contextualization, comparison, and collaboration.
% Reproducibility
To keep track of the analysis workflow, we have based the design of Immersive Insights on the concept of a ``data instance'' (also referred to as a ``solution''): any unique combination of data samples and dimensions that originates from the same, original dataset - and onto which parametric algorithms can be independently applied. Our definition of a ``data instance'' is analogous to the concept of the ``clustering view'' proposed in \cite{cavallo2018clustrophile}, where the authors mention screen size as a limiting factor to the number of data instances that can be concurrently analyzed.
%AR/VR
By redefining classic EDA workflows within the virtuality spectrum, and by leveraging a hybrid reality system with about 20 times the resolution and 100 times the screen surface of a laptop, we attempt to provide a tool with better contextualization and comparison capabilities, supporting simultaneous independent analyses on multiple data instances.
%Collaboration
To keep track of data instances and foster collaborative analysis, we further combine Dataspace screens with AR devices and the central table, which offers a summary view of the entire session.

\subsection{Data Views}

\begin{figure*}
\includegraphics[width=\textwidth]{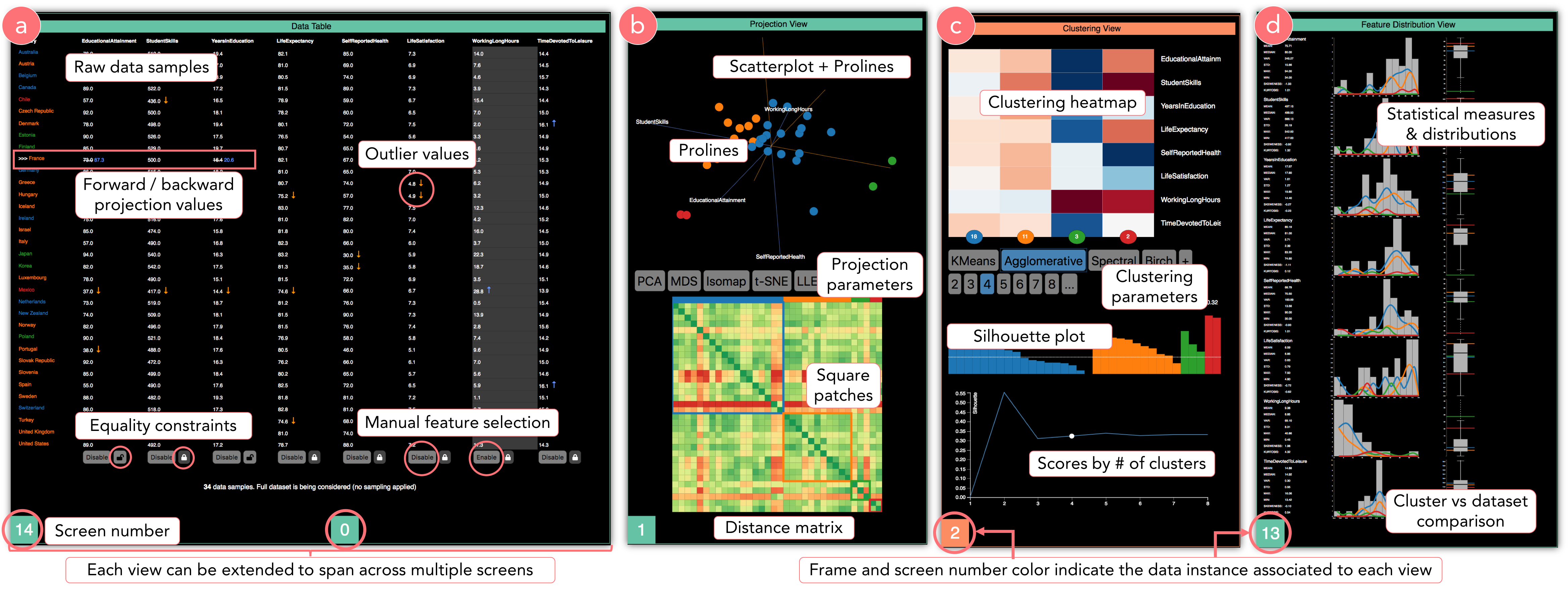}
\vspace{-1em}
\caption{\textbf{Sample data views}. Detailed statistical information on data instances is visualized in ``data views,'' graphical layouts that can be dynamically loaded onto the Dataspace screens. Each data view is associated with a specific data instance or solution, and can be extended to multiple screens. The figure above shows a) a data table view, b) a projection view, c) a clustering view, and d) a distribution view. Please refer to the supplemental video for a more complete picture of available data views.}
\label{fig:views}
\vspace{-1.5em}
\end{figure*}

%Definition
Visualizations, statistical information, clustering and projection associated with a specific data instance are grouped into dedicated graphical layouts, called ``data views''.
%How to access them
Data views can be loaded onto Dataspace screens through voice interaction (e.g. by saying ``Show projection view on that screen / on screen number 13''); from the AR headsets; or from a touch menu that can be opened on the table.
Whenever a data view is loaded, the robotic arm associated with the screen moves down to bring the screen closer to the user.
%Similarly, views can be cleared and associated screens pulled back up.
%Expanding view
By default, each view spans one Dataspace screen, but can be arbitrarily extended to occupy multiple screens to accommodate larger amounts of information (e.g. by saying things like ``Load X view on screens Y and Z'' or ``Extend X to N screens'').
% Grouping
Each data view is bound to a specific data instance (solution), each of which can be identified by the frame color of the screen. Thanks to this functionality, the same type of data view can be independently used on different screens and bound to different data instances.
%List the views
Despite that the data views implemented in our system largely make use of existing visualizations and interaction types \cite{becker1987brushing,buja1991interactive,wills2008linked}, we still find it useful to discuss the rationale behind their design in the context of a hybrid reality environment and as a base for understanding the user study proposed in Section 4.
%Also note design choices such as the use of black backgrounds to reduce energy consumption and overheating, the large size and positioning of buttons to facilitate touch, and separate color palettes encoding clusters (saturated) and data instances (pastel).

\subsubsection{Data table (Fig.~\ref{fig:views}a)}
It is important to provide users with a simple way to access and form initial intuitions about raw data---down to the scale of a single data sample.
The Data Table view augments the visualization of data samples with a simple form of outlier detection, and uses color to encode eventual clustering assignments. Touch and voice interactions can be used to select and sort data samples, and to ``enable or disable'' data features (i.e. consider them or not in the analysis).
The Data table also plays an important role in the process of forward and backward projection, explained in Section 3.4. We note that, despite that this view can be easily extended to multiple screens to show more data features, it still does not scale to high-dimensional datasets \cite{gratzl2013lineup}---and this is the reason why we included an option to show only the most relevant dimensions identified in the \textit{feature selection view}.

\subsubsection{Projection (Fig.~\ref{fig:views}b)}
Dimensionality reduction algorithms are a convenient way data scientists visualize high-dimensional data.
% Structures and outliers
Dataspace's Projection view enables users to specify a dimensionality reduction algorithm and a varying number of parameters (e.g. distance metric) and visualize the algorithm’s output as a 2D scatterplot, where each data point is colored according to its clustering assignment.
% Feature interpretability
While a very powerful means to identify structures and outliers in the data, scatterplots of dimensionally reduced data generally lack interpretability as to the contribution of specific data features to the projection. To mitigate this, we complement the scatterplot visualization with \textit{prolines} \cite{cavallo2018visual}, a generalized version of biplot \cite{gabriel1971biplot} that introduces axes representative of the original data dimensions. Each proline axis indicates the relevance and directionality of increase for a feature, and provides statistical information about that feature’s distribution.
%[C] through dedicated glyphs.
% Debugging
In order to allow further debugging of dimensionality reduction algorithms and, more specifically, of their distance metrics, the Projection view includes a heatmap indicating per-point distances, where rows and columns are sorted by clustering assignments. In the case of good cluster compactness and separation, the heatmap clearly visualizes different ``square color patches'' along its diagonal. Uniform colors across the distance matrix may indicate instead points that are too far apart from each other in multi-dimensional space (``curse of dimensionality''), requiring a re-definition of the adopted distance metric.
% Interactions
Scatterplot points and matrix cells in the Projection view are linked to all other data views associated with the same data instance, enabling the visualization of contextual information after performing a selection on any view.

\subsubsection{Clustering (Fig.~\ref{fig:views}c)}
% CLustering
Clustering algorithms are generally used by data scientists to identify groups of data points that have similar features, dividing the dataset into a number of clusters (often user-defined). Each clustering algorithm depends on a varying set of parameters (such as distance metric and number of clusters to apply), and outputs a per-point cluster assignment (``label'') or a probability of belonging to a certain cluster. In Immersive Insights, these algorithmic choices can be expressed through touch from the Clustering view or voice (e.g. by saying, ``Apply agglomerative clustering with 4 clusters to solution 1''), and clustering results are broadcasted to all views associated with the current data instance.
% Interpretation
To facilitate the interpretation of clustering assignments with respect to cluster sizes and statistical properties of each cluster, Immersive Insights' Clustering view includes a heatmap visualization where columns represent clusters and rows represent data features. The color of each matrix cell encodes the average feature value for a particular cluster, with red being very high and blue very low. By looking at a column vertically, users can quickly spot key features characterizing a cluster; by looking at a specific row, users can compare the values of multiple clusters based on that feature.
%Silhouette plot
A fundamental step in clustering analysis involves evaluating the ``goodness'' of a clustering result, which is generally quantified in terms of cluster \textit{compactness} 
%[C](how close members are to each other)
 and \textit{isolation}
%[C] (how distant members are from other clusters) 
\cite{liu2010understanding}. The Clustering view displays silhouette score \cite{rousseeuw1987silhouettes} as a validation metric, and includes a silhouette plot to help scientists understand which data points will likely change a clustering assignment
%, and which clusters will likely merge with others 
when clustering parameters are changed. A line chart with precomputed silhouette scores further enables users to intuit how cluster separation and compactness vary based on the choice of number of clusters to consider.

\subsubsection{Distribution (Fig.~\ref{fig:views}d)}
Statistical information plays a fundamental role in EDA, and visualizing the distribution of each feature helps data scientists understand the properties of their data, as well as debug the outcomes of applying various algorithms (e.g. normalization, skewness of data). Therefore, our distribution view includes distributional information on features, in the form of summary statistics, histograms and box plots.
In particular, we use colored overlays and side-to-side summaries to enable statistical comparison of identified structures (selections, clusters) with respect to the full-dataset distribution. This allows users to intuit the size of each cluster, its differentiating features, and its impact on determining the global data distribution.

\subsubsection{Correlation}
Identifying relevant correlations among data dimensions is another process fundamental to EDA. On top of indicating possible predictive relationships among features, correlations are also used to generate initial hypotheses about the data, and for feature selection. In the presence of multiple variables that are highly correlated, for instance, data scientists may work to remove noise by electing to keep only one such variable, or by substituting in a new feature synthesized from  the original ones. Our correlation view displays a bar chart with top pairwise correlations, and a scatterplot matrix of pairwise correlations where each data point is colored based on its clustering assignment.
%so users may examine feature distributions in greater detail, and in the context of clustering information (encoded by the color assigned to each scatterplot data point).

\subsubsection{Feature selection}
Feature selection is the process of deciding which of the original data dimensions should be included in an analysis, or fed into an algorithm. This process aims at improving computational performance during analysis, removing noise data, or improving estimators' accuracy scores on high-dimensional datasets. We consider two types of feature selection: algorithms that can be applied to fully unsupervised data (e.g. PCA, variance, feature agglomeration) and algorithms that rely on labels generated during the clustering process (e.g. univariate selection with chi squared or ANOVA). In the former case, features are ranked based on variation in their distributional information; in the latter, features are ranked based on their relevance (p-value, effect size) in determining a specific clustering outcome. Users can manually decide to enable or disable features or allow the system to automatically select the most \textit{n} relevant ones.

\subsubsection{Feature filtering and engineering}
%Feature filtering
Successful EDA requires the ability to consider meaningful selections from the original data, and alternatively focus analysis on these subsets. For this reason, we include a filtering feature that allows users to select data samples based on AND/OR combinations of different equality and inequality constraints. These constraints are defined through touch, voice, or keyboard input. Once a selection is made or a filter applied, the chosen points are highlighted in all views associated with this data instance. The user may also perform \textit{isolation} and \textit{reprojection}, which, respectively, hide all non-selected points from all views, and update scatterplots of dimensionally reduced data by recomputing their projections only on the selected samples.
%This process is particularly effective in safely and iteratively analyzing small structures  independent of the rest of the data.\\
%Feature engineering
Immersive Insights further allows users to add new features to the dataset that are  generated as combinations of existing features, a process often referred to as ``feature engineering''.
%Feature engineering is particularly useful in the presence of highly correlated features or when performing feature agglomeration. Therefore, '

%[C]\subsubsection{Database selection}
%[C]The database selection view, automatically loaded when Immersive Insights runs, enables users to import or load recent data, to specify data sampling rate and default data view configurations, and to receive visual feedback on the correctness of voice commands.

\subsection{Table View: Analysis Overview}
\begin{figure}
\includegraphics[width=\columnwidth]{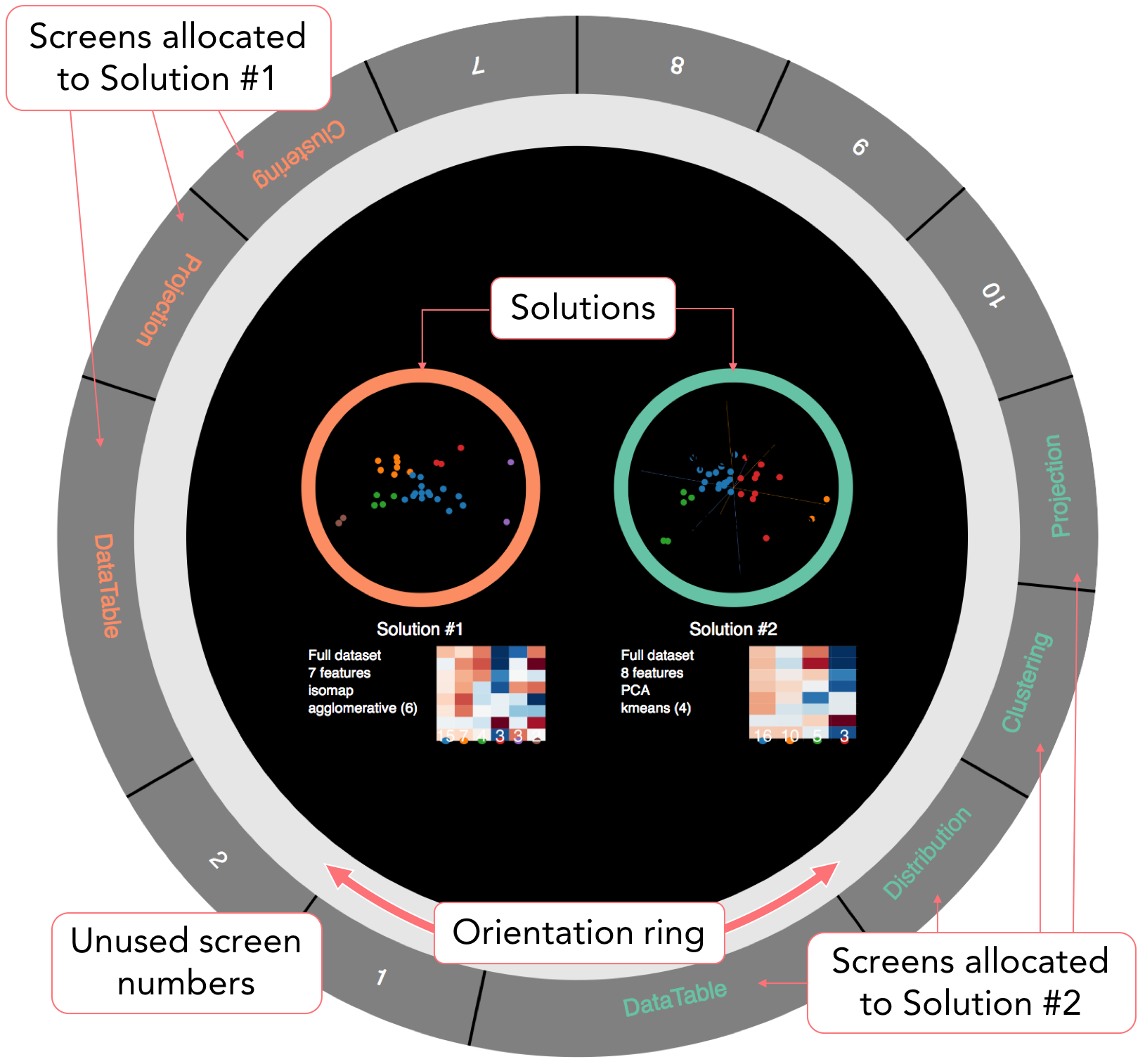}  \caption{\textbf{Table view}. Visual content is projected onto the central table, providing an overview of the current state of the analysis and a list of solutions identified by data scientists. To help users locate content in the environment, an outer ring displays which data views have been loaded onto Dataspace screens.
% and with which solution they are associated. 
Touch interaction can be used to rotate table content towards a particular user through the ``orientation ring'' and to manage data views.}
\vspace{-1em}
\label{fig:table}
\end{figure}

Managing multiple data views and tracking the analysis workflow are non-trivial tasks whose complexity increases in collaborative settings. Immersive Insights allows data scientists access to their own slice of data for personalized analysis through the concept of \textit{data instances} (solutions), requiring them to keep track of which Dataspace screens and which parameters are associated with each solution. At the same time, it is important to have a central gathering point where users can discuss their opinions on the use of algorithms or on the labeling of data points, facilitating consensus or more efficiently incorporating ideas from alternative procedures.
% Table solution
With these considerations in mind, we included a circular view on the central projection table, with the two-fold purpose of orientating users in the environment and enabling comparison of different analysis solutions (Fig.~\ref{fig:table}).
% Representing solutions
We visualize each data instance by combining the dimensionality reduction 2D scatterplot and the clustering heatmap visualization, which together represent the minimum amount of visual information necessary to qualitatively describe a clustering result \cite{cavallo2018clustrophile}. Textual information about the algorithms generating each solution and their silhouette scores are also shown, providing a simple form of quantitative comparison.
% Relation with AR view
By tapping on a particular solution, users can enable or disable the corresponding 3D projection in the AR view, allowing  for side-to-side spatial comparison.
% Associating views
The Dataspace screen(s) associated with a particular solution can instead be identified by 1) visually matching the color of the solution on the table with the color of the frame of particular Dataspace screen(s), 2) by double-tapping a solution on the table, an action that highlights associated data views and physically brings their screens closer to the table,
%[C] using the robotic arms
 and 3) by referring to the colors displayed in the outer ring of the Table view.
% Rings
In fact, we have included two interactive rings on the table. The outer ring is divided into small sections of variable radius, each representing a data view and its associated Dataspace screen(s). A user studying the data view on the table in front of him/her can simply look up and find that same view on the screen(s) ahead.
%Each section on the table is color-coded based on the solution associated with that data view, and each is proportionally sized  based on the number of screens employed by that view.
When comparing results at the table and preparing for the next steps in the analysis, it is often necessary to reorganize data views. The outer ring supports drag-and-drop interactions, allowing users to move a view to a different screen, extend it to multiple screens, or clear that view.
The inner ``orientation'' ring, on the other hand, is used to rotate the central content of the Table view towards a particular user \cite{Kruger:2004:ROT:1045432.1045437}.

\subsection{Augmented Reality Integration}

Immersive Insights supports the optional use of augmented reality headsets (Microsoft Hololens and Magic Leap One) to complement the data analysis experience. AR devices are used to visualize data atop the central table, but also to provide guidance and information on-demand when interacting with specific Dataspace views. Thanks to the environment's spatial awareness system, AR-specific interaction such as the use of gaze and airtap (Hololens) and 3D controllers (Magic Leap One) are transformed through raycasting into touch-equivalent events, enabling AR users to perform standard operations on Dataspace screens from a distance.

\paragraph{Enhanced exploration of high-dimensional data} Thanks to their ability to visually summarize a data instance immediately upon being looked at, projection views of dimensionally reduced data (Section 3.2.2) are arguably the most used type of view in Immersive Insights. While they provide users with quick, qualitative intuitions regarding the effects of any algorithm applied to the data, they also generally require the user to turn back and forth in order to look at other complementary views (e.g. clustering, distribution) and quantify the results. Hence, in Immersive Insights we propose a tridimensional, AR version of the Dataspace projection view, to be visualized atop the central table.
This makes the view simultaneously visible to all AR users from any location, and allows a seamless transition from AR content to detailed statistical information on the environment's screens.
%[C] (without requiring users to remove the headset).
% 3D
The ability to visualize an extra dimension in the dimensionality reduction, and the larger surface (volume) available for the projection, together represent an opportunity to improve the interpretability of projection axes, reduce visual clutter, and introduce direct manipulation of data samples.
% Perspective, pan & zoom
%. Prolines are now defined in three dimensions as well. Due to users being able to dynamically select  a particular viewing perspective on the data, these prolines encourage  a better understanding of feature importance and cluster separation.
% Forward and backward projection
To support sensitivity and what-if analysis, we have introduced a tridimensional generalization of the forward and backward projection interactions originally proposed by Cavallo in \cite{cavallo2018visual}.
Through \textit{forward projection}, the user can specify through voice (or manually from the Data table view) a perturbation in the feature values of a selected data sample (e.g. ``Try increasing the <feature> value of this data point by 5''), and observe how the position of the point changes via an animation displayed in the AR projection view. Conversely, a data scientist can use \textit{backward projection} by manually dragging a data point around in the projection space, and observing how its feature values would have to change to accommodate the introduced modification.
For example, a user might consider an outlier data point, questioning how that point's feature values would have to change in order for it to belong to the neighboring cluster. Simply dragging the point closer to that cluster helps quantify how much each feature has determined the point's distance from the cluster.
We refer readers to the original paper \cite{cavallo2018visual} for more details on prolines and forward and backward projections.
%on the guidance provided by prolines during forward and backward projections and on the definition of constraints in backward projection (which in our case can be specified from the Data table view).

\paragraph{Providing guidance during the EDA workflow}
On top of using it for a dedicated central projection view, we leverage AR to provide further visual assistance to users during interactions with standard Dataspace views.
% Screen highlighting
When the wearer of an AR headset looks at (Hololens) or orients their controller towards (Magic Leap One) a data view, an AR overlay frame is created around the view and a cursor is rendered on the physical screen through raycasting, clarifying which UI elements the user is interacting with. This feature is particularly useful for managing tracking or alignment issues associated with the relationship between the AR view and the physical environment.
Similarly, when an action is performed on the data, all views affected by this change are momentarily highlighted in the AR view and virtual arrows are used to indicate updates in views situated behind the user.
% Voice
Similarly, speech transcription in AR allows for real-time feedback when using voice commands, providing a way to debug unsuccessful commands and minimizing the time it takes to correct them (e.g. in the case of incorrectly spelling an algorithm's name).
% Data view-specific
Finally, AR can be used to provide on-demand information specific to each data view. Due to screen space and readability limitations, it is unreasonable to permanently add more information (such as suggestions or feedback on the choice of algorithms) to existing data views. However, augmented reality allows us to dynamically display additional information without modifying the layout of existing views, and personalize that information for the user who currently needs it. Current implementations of this feature, activated by user proximity to a screen, include bubble overlays with statistical information (e.g. area distribution charts in the Feature selection view, ANOVA p-values in the Clustering view) and simple virtual UIs to facilitate certain tasks (e.g. a slider to modify the number of clusters in the Clustering view).
% + suggestions

\section{Evaluation}

\begin{figure}
\includegraphics[width=0.82\columnwidth]{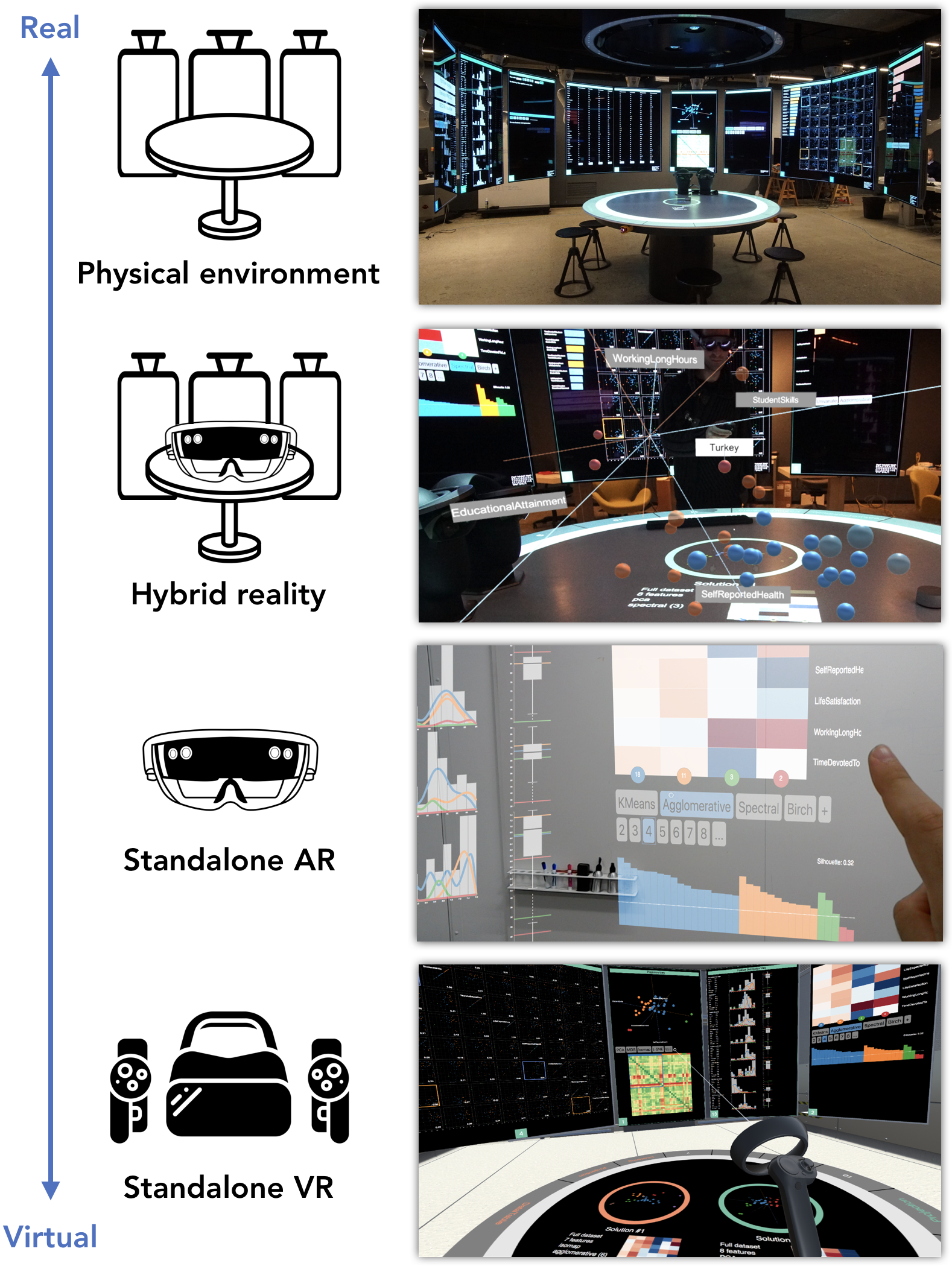}
\caption{\textbf{Virtuality continuum}. In the first part of our user study, we evaluated individual task performance using Immersive Insights in four different modalities. In the standalone AR and VR modalities, the physical environment was replaced by a virtual replica of Dataspace.}
\label{fig:continuum}
\vspace{-1em}
\end{figure}

We illustrate here the procedure and results of an initial, two-part user study conducted on 12 data scientists using Immersive Insights.
%1st part
The goal of the first part of the study was to evaluate the contributions and limitations of immersive technologies in EDA at various levels of the \textit{virtuality continuum} \cite{milgram1994taxonomy}, leveraging the flexible data immersion capability of Dataspace and its virtual model \cite{cavallo2019dataspace}. Specifically, we had participants perform very specific data analysis tasks with Immersive Insights in four different Dataspace modalities (Fig.~\ref{fig:continuum}): 1) without AR integration, 2) with AR integration, 3) using a virtual representation of Dataspace in AR (i.e. using virtual screens in place of the physical ones), and 4) fully immersed in VR.
%2nd part
In the second part of the study, our goal was to evaluate the effectiveness of Immersive Insights in allowing data scientists to collaboratively generate insights, comparing our system with the recent desktop-based tool \textit{Clustrophile 2}. As opposed to the first part of the study, here we focused on the data analysis session as a whole, allowing participants complete freedom while collaboratively solving a single, complex task.
% starting with an empty Dataspace configuration (i.e. with only the Data Table view preloaded).
We compared Immersive Insights with Clustrophile 2 for two main reasons: 1) Clustrophile 2 is a fully-implemented system that extends beyond the scope of a simple research prototype, allowing for an end-to-end analysis session; and 2) despite differences among technologies employed, the two systems share a similar design and present almost equivalent functionality.
% and 3) the user study methodology we elected to use is similar to that proposed by Cavallo and Demiralp \cite{cavallo2018clustrophile}.
%End
Finally, we conclude by analyzing our results and provide some considerations regarding the use of immersive technologies for collaborative data analysis.

\paragraph{Participants.}
We recruited a total of twelve data scientists, four for each of the three data analyst archetypes (hacker, scripter, and application user) identified by Kendel et al. \cite{kandel2012enterprise}. In addition to ensuring full coverage of different types of users, this choice of participants made it easier to compare the performance of our system to the results obtained in the Clustrophile 2 user study. Participants were selected from among 20
candidates who had at least a masters degree in science or engineering and at least two years of work in data science. Selections were made so as to maintain gender parity, with ages ranging from 25 to 43 years old. 
%We excluded any participant who might have previously analyzed any of the datasets involved in our study.

\paragraph{Procedure.} Participants were shown a 10-minute video on user interactions in Dataspace, then individually introduced to the physical environment. Assisted by one of our collaborators, participants were then invited to freely analyze a toy dataset focused on indices of wellness for 34 OECD countries \cite{oecd}.
% using both Immersive Insights and Clustrophile 2.
Initially, each participant was required to \textit{individually} perform the same data analysis tasks according to only one of the four possible modalities (i.e. three participants per each modality in Fig.~\ref{fig:continuum}).
In the second part of the study, the same participants were placed in groups of three (one hacker, one scripter and one application user per group), so they could collaborate on a \textit{team} solution.
We performed controlled assignments to balance the distribution of user archetypes and the use of AR and VR headsets.
We recorded videos of both parts of the study, asking each group to think through their actions and interactions with teammates aloud. During the user study, analysts were not permitted to ask our collaborators about anything beyond the usage of a particular feature of the system.

\subsection{Part 1: EDA and the Virtuality Continuum}

The first part of the study involved \textit{individually} performing four EDA tasks, comparing their time performance by using Immersive Insights in four different modalities:
\begin{itemize}
\item Real environment: only Dataspace screens and table, and their associated touch, gesture and voice interactions, could be used by participants.
\item Hybrid reality: AR headsets could be used together with Dataspace physical elements, as described in Section 3.
\item Standalone AR: participants wore a Hololens headset and interacted only with a virtual replica of Dataspace (1:1 scale), originally developed by the authors to support remote participation \cite{cavallo2019dataspace}. In this setting, all Immersive Insights functionalities were preserved, but physical screens and table were replaced by digital counterparts, and touch interactions were taken over by raycasting and airtap gestures. Voice commands were still enabled thanks to the device's microphone.
\item Standalone VR: similar to the AR standalone setting, participants used a virtual replica of the Dataspace environment, but were completely immersed in VR through a Samsung Odyssey headset. The usual touch interactions were replaced by raycasting through the Samsung Odyssey controllers, who could also be used to move around the virtual environment using the standard teleportation feature.
\end{itemize}
%The data
The dataset used for the experiment consisted of 200 samples randomly chosen from a larger database of Fitbit activity recordings. Each row was associated with the activity levels of single subject over a period of three months, and was characterized by 19 data dimensions, including average number of steps during weekdays and weekends, gender and age information, and other indices such as stress and wellness score.
%The procedure
Each participant performed tasks independently, and, with the exclusion of the first modality, was required to wear the headset (Hololens or Samsung Odyssey) from the beginning until the end of the experiment. In all four modalities, participants started each task with the same initial Dataspace configuration, composed of a total of 10 Dataspace screens displaying each of Immersive Insights' data views.
%TODO: A task would be considered unsuccessful if
%Tasks
The tasks, selected to be non-trivial and to involve the use of multiple data views, were the following:
\begin{itemize}

\item \textbf{T1}: Identify the person with the highest participation rate among the cluster of subjects who are mostly active on weekdays. Compare that person's stress and wellness scores with those of the other members of the cluster.
\item \textbf{T2}: Explain how age and gender affect preferences regarding days on which to engage in physical activity among the cluster of less active subjects.
\item \textbf{T3}: Find out whether, among the group of most active participants, people who engage in activity on weekdays do so more regularly than those who prefer to go running on weekends.
\item \textbf{T4}: Considering that the subject with ID 1036 cannot currently change his workout schedule, suggest and quantify what this participant should do to improve his overall wellness score without increasing his stress level.

\end{itemize}

\subsection{Part 2: Hybrid Reality vs Desktop-based}
We compared Immersive Insights to Clustrophile 2 by using the same dataset and task proposed by Cavallo and Demiralp in their user study \cite{cavallo2018clustrophile}. Their study design specifically aims at evaluating a real-world, unconstrained EDA session, where participants are asked to answer a single, open-ended research question. Since no univocal solution exists, participants can decide when they have reached a satisfiable result, ending the session.
% Add why we used the same design
%The data
The adopted dataset has 8652 rows and 37 features, and is generated by preprocessing patient data made publicly available by the Parkinson's Progression Markers Initiative (PPMI). Specifically, this dataset contains UPDRS (Unified Parkinson's Disease Rating Scale) scores, which consist of a set of clinical measures describing the severity of each individual's motor condition.
%The procedure
We assigned to each of the four groups of data scientists the task of identifying plausible Parkinson's phenotypes and mentioning aloud during the analysis all relevant insights they gained about the data. Each group started with a data table view spanning two screens, and two Hololens devices available on the table.
%One volunteer from each group was required to wear a Hololens or Magic Leap One from the start.
%Comparison
%To evaluate the results, we compared the insights \textit{collaboratively} arrived at  by each team using Immersive Insights with cumulative insights individually collected by Clustrophile 2 users, grouped according to analyst archetype.
Since Clustrophile 2 does not support collaborative analysis, we grouped its users by analyst archetype as to match the groups who used Immersive Insights. This way, we were able to compare Immersive Insights' \textit{collaborative} results with the cumulative insights \textit{individually} collected by Clustrophile 2 users.
We note that the list of insights shown in Fig.~\ref{fig:results} was manually composed by the experimenter at the end of the user study by combining findings (collected through our think-aloud protocol) from the different sessions. The list does not include false insights.

\subsection{Results and Discussion}

\begin{figure}
\includegraphics[width=\columnwidth]{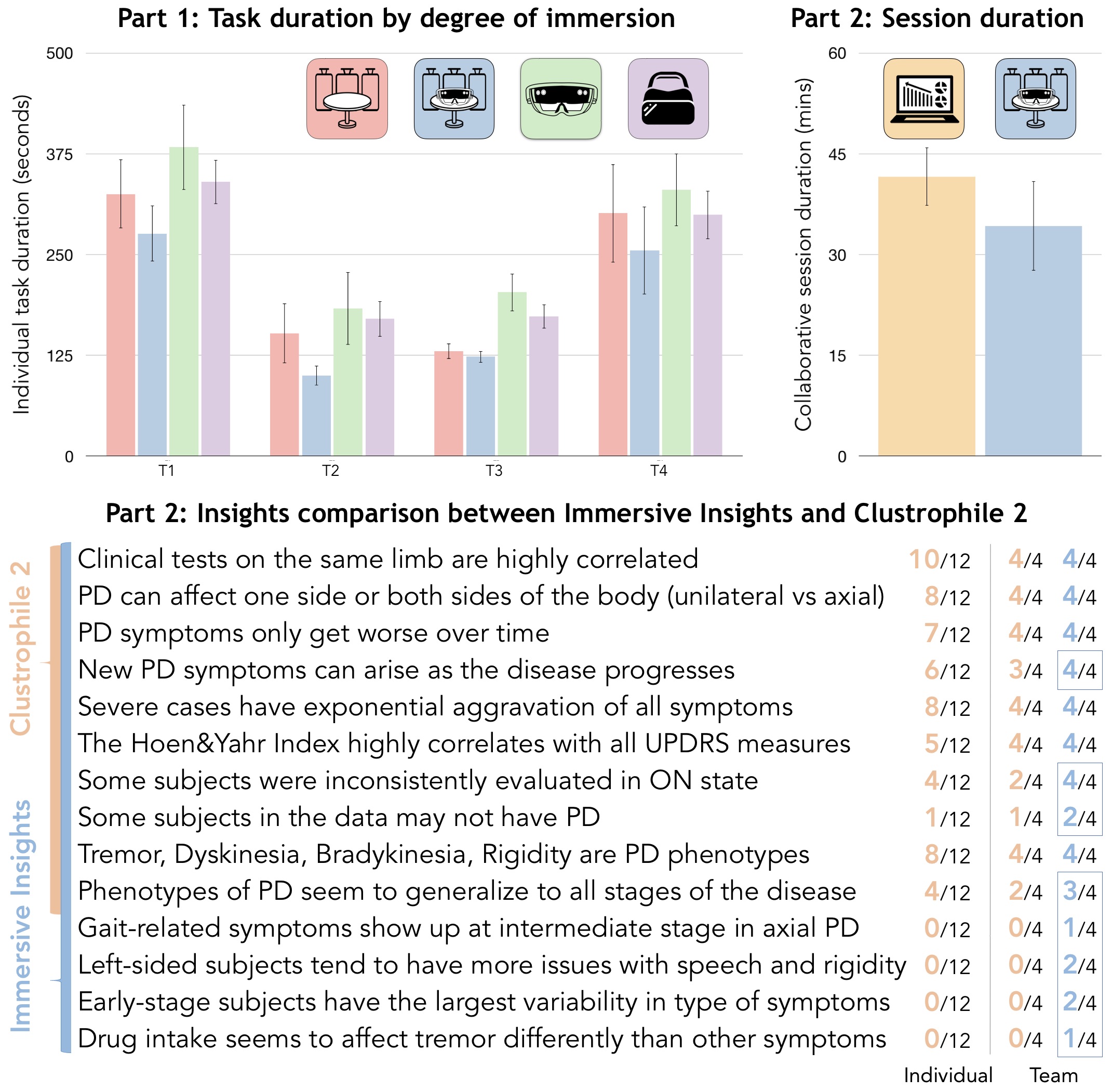}
\caption{\textbf{User study results}. The first part of the user study suggested that the integration of AR in EDA workflows can improve analysis time for certain tasks. The second part showed how collaborative analysis through Immersive Insights allowed quicker arrival at insights than did a more classic analysis in Clustrophile 2.
%We note that 
Insights shown here are not meant to be interpreted as medical evidence.
%Error bars shown the bar plots indicate the standard error of each distribution.
}
\label{fig:results}
\vspace{-1em}
\end{figure}

Results of the two parts of the study are summarized in Fig.~\ref{fig:results}.
% 1st
%Integrating augmented reality into the EDA workflow led to an improvement in task duration. However, results also demonstrated that standalone AR and VR underperformed with respect to our hybrid reality solution, proving that immersive technologies still have a long way to go before entirely disrupting standard EDA methodologies.
While the number of participants limits the statistical significance of these initial findings, our results suggest that integrating augmented reality into the EDA workflow can lead to an improvement in task duration. However, results also show that standalone AR and VR underperformed with respect to our hybrid reality solution, indicating that immersive technologies still have a long way to go before entirely disrupting standard EDA methodologies.
% 2st
Immersive Insights users were able to arrive at more insights, and in less time, than Clustrophile 2 users, suggesting that the development of hybrid analytics tools would be beneficial for EDA.
Below, we report some considerations and statistics based on direct observation of the participants and post-experiment verbal feedback.

\subsubsection{(Confirmed) Limitations and Challenges}

While various examples from the literature show that AR and VR headsets can be effectively used to perform specific data analysis tasks \cite{millais2018exploring,butscher2018clusters,marriot2018immersive}, durations recorded in the first part of our study indicate that recent technological advances in interaction, resolution, and field of view of HMD devices are still limiting factors in the application of these technologies to EDA.
%%%AR
% Interaction
In the case of standalone AR, the Immersive Insights modality that scored the lowest performance, task duration was considerably affected by interaction with data views through the Hololens airtap gesture and the gaze-based pointing technique. In particular, airtap proved to involve a steep learning curve and, despite the enlarged size of UI affordances, gaze tracking was imprecise when interacting with views from the middle-to-long distance (57\% successful airtap attempts overall). We expect this accuracy to improve with the use of the Hololens \textit{clicker} device, which was not available in this study.
% Field of view and resolution
Limited field of view, a known issue with Microsoft Hololens, played a crucial role in the standalone AR modality, forcing users to frequently rotate their heads and therefore completely lose the advantages of peripheral vision in combining contextual information from multiple (virtual) views. In most cases, participants were required to move very close to various views in order to see tables and statistical information, which were otherwise difficult to interpret due to the device's resolution.
% VR
Standalone VR overall had a better performance than AR, mostly due to the use of hand controllers, the higher resolution and field of view of the Samsung Odyssey, and to the possibility of teleporting in the virtual room (e.g. to next to the virtual screens of interest). Fine-grained UI control through raycasting, however, still proved unreliable with respect to touch interaction, and the HMD's pixel-density could not match the high resolution statistical info displayed by physical Dataspace screens.\\
% USability
The second part of the study, which involved a more prolonged use of these devices, further confirmed discomfort in wearing headsets (group average of 8.1 minutes of intermittent use,  about one fourth of the average session duration), gesture interpretation issues (e.g. an involuntary \textit{bloom} gesture shut down the Immersive Insights application on two headsets), and difficulties with text input (e.g. users fell back to using the Dataspace mechanical keyboard for filtering operations).
% Conclusion
Though their use is tempting, we should not yet conclude that modern HMDs and virtual replicas are ready to render more expensive technologies such as physical collaborative environments obsolete.

\subsubsection{Collaborative Hybrid Analytics}
Immersive technologies promise to enrich EDA through spatial data analysis, on-demand statistical information, more flexible management of contextual content, and (remote) collaboration. While our user study reported unclear advantages in utilizing depth information while visualizing high dimensional data \cite{bach2017hologram}, the comparison of Immersive Insights with Clustrophile 2 demonstrated the potential contributions of a \textit{hybrid reality} approach to collaborative data analysis.
By summing the number of parameters applied by each group during a session (e.g. the number of changes in algorithm, metrics, cluster number, and projection), we noted that Immersive Insights users attempted less parameter changes than Clustrophile 2 users (19.7 vs 27.74 average changes), and performed more simultaneous independent analyses (2.7 vs 1.83 average data instances), with an overall 21.6\% improvement in total time spent. We believe these results are connected with 1) the intrinsically collaborative setting of Dataspace (e.g. group discussion limited the trial-and-error approaches seen with Clustrophile 2), 2) the Table view (e.g. comparing and keeping track of data instances), and 3) the enhanced ability to contextualize provided by the larger total screen surface and seamless integration with AR augmentations.
An average of 1.4 devices were used by each team during the analysis session, with men wearing the headset 2.2 times longer than female participants. We also noted an interesting tendency to pass a device to another member of the team, rather than picking up new devices from the table.
%To improve the statistical significance of our results, and to further our understanding of team dynamics in collaborative hybrid analytics, we plan on conducting a more extensive future study with more participants. The study will also include an evaluation of how remote collaboration through AR and VR may impact the EDA process.
To further our understanding of team dynamics in collaborative hybrid analytics, we plan on conducting a more extensive future study, which will also include an evaluation of how remote collaboration through AR and VR may impact the EDA process.
%``I think one of the main advantages [of Immersive Insights] is that I can easily contextualize the data I am focusing on without having to continuously change tabs on my laptop'', commented one participant.
%Interactions

%having a central projection was useful in observing the data, and that it minimized the time it took to change focus from data views and table to the AR view, and vice-versa.
%we noticed that people who started the analysis wearing an AR device where more likely to keep it on until for longer time.
%In particular, results showed less total interactions with the system and more statistical exploration of substructures through filtering and isolation.

\subsubsection{The Need for Flexible Data Immersion}
% Too much hype
The new wave of AR and VR devices has undoubtedly generated excitement about their applicability to many domains, including EDA, and their use has the potential to benefit various aspects of data analysis.
% The reality
%However, if we exclude developers and AR/VR enthusiasts, our study demonstrates that most people cannot spend more than few minutes wearing a Hololens.
However, our study confirmed that hardware and comfort limitations still inhibit the use of Hololens for extended periods of time during the analysis.
We also observed that standalone AR/VR experiences are not suited for a fully-fledged EDA, as many design and interaction challenges still need to be addressed.
In fact, while voice commands, wand-based selections, and 3D manipulations can be useful for certain visual analytic tasks, a 2D monitor with touch or mechanical keyboard input remains significantly more efficient for other tasks.
% Suggestion
While our community continues researching appropriate ways to apply these technologies to data science, we believe AR and VR should be cautiously used to \textit{complement}  existing visual analytics methodologies, rather than imagined as a possible replacement.
In particular, we encourage the development of systems that offer multimodal interaction and a \textit{flexible level of data immersion}, so that users are free to decide at any moment which technology and interactions makes sense to use when performing specific EDA tasks.

\section{Conclusion}

In this work, we presented Immersive Insights, a \textit{hybrid analytics} system for exploratory data analysis (EDA) implemented in a Dataspace collaborative environment.
Leveraging the flexible degree of immersion provided by Dataspace through its integration with AR and VR devices, we were able to study how these rising technologies can contribute to EDA workflows at different levels of the \textit{virtuality spectrum}. We also evaluated Immersive Insights by comparing its performance with a similar desktop-based tool during a collaborative analysis session.
While we acknowledged that immersive technologies can contribute to reduced analysis time and may facilitate the generation of additional insights, we also demonstrated that AR and VR cannot yet entirely replace non-immersive EDA methodologies.
Ultimately, we encourage the development of hybrid systems in which immersive technologies do not aim to take over existing EDA workflows, but try to complement them instead.

\balance

% Bibliography
\bibliographystyle{ACM-Reference-Format}
\bibliography{bibliography}

\end{document}